\documentclass[epj]{svjour}
\usepackage{graphicx}

\begin{document}
\title{Crack patterns in drying protein solution drops}
\author{C.C. Annarelli, J. Fornazero, J. Bert and J. Colombani}
\institute{D\'epartement de Physique des Mat\'eriaux - B\^atiment Brillouin\\
Universit\'e Claude Bernard Lyon 1\\
6, rue Amp\`ere, F-69622 Lyon-Villeurbanne cedex, France}
\titlerunning{Crack patterns in drying protein solution drops}
\authorrunning{C.C. Annarelli \textit{et al.}}
\abstract{A deposited drop of bovine serum albumin salt solution experiences
both gelation and fracturation during evaporation.
The cracks appearing at the edge of the gelling drop are regularly spaced,
due to the competition between the evaporation-induced and
relaxation-induced stress evolution.
Subsequently, the mean crack spacing evolves in an unexpected way, being
inversely proportional instead of proportional to the deposit thickness.
This evolution has been ascribed to the change with time of the average
shrinkage stress, the crack patterning being purely elastic instead of
evaporation-controlled.}
\mail{Jean Colombani\\
(Jean.Colombani@dpm.univ-lyon1.fr)}
\maketitle

\section{Introduction}
The study of crack dynamics in drying materials meets the basic interest
in nonequilibrium pattern formation and the industrial need of controlling
drying processes (paint technology, \dots).
But whereas the understanding of crack behaviours is growing, partly due
from the interplay between mechanics and statistical physics approaches
\cite{Charmet}, the investigation of the cracking of drying complex fluids
is just setting in.

As part of a general study of biocompatibility between prosthetic materials
and real biological liquids, we focussed on the evaporation of a protein
solution drop and on the evolution of the solid deposit.
Although the gelation process of our biological liquid and the initial
stage of the crack pattern formation show similarities with like phenomena
in mineral colloids \cite{ParisseL,Pauchard}, the characteristics of our
experiments yield further insight in the following steps of the cracking.

Our working liquid is a phosphate buffered saline solution ($pH=7.4$, ionic
strength $I=0.2$ M) of bovine serum albumin (BSA) at various
concentrations, which may be classified as a colloidal sol with the BSA molecules
amounting to ellipsoidal particles of dimensions $4\times 4\times 14$ nm$^3$
\cite{Peters}.
The volume fraction of protein is always a few percents in our
concentration range.
In the course of the experiment, a $15$ mm$^3$ sessile drop is deposited onto
a glass slide.
Great care is taken for the cleaning of the substrate, residual deposits being
liable to disturb the wetting and cracking behaviour.
Keeping in the surface tension-driven regime, tensile drops have been
experienced to exhibit the same behaviour.
Immediately, proteins are adsorbed on the substrate, leading to a
strong anchoring of the triple line \cite{AnnarelliJCIS}.

The first step of dessication is then characterized by a fixed base radius
of the drop, a constant evaporation rate all over its free surface and a
spherical cap shape induced by surface tension (Bond number smaller than one).
These three features have been shown to remain compatible thanks to outward
flows inside the drop \cite{Deegan}.
Subsequently an apparently solid ring-like ``foot'' progressively forms close to
the edge of the drop, and the three phase line migrates inwards, provoking the
extension of the solid until complete dessication.

Due to the competition between the adhesion of the deposit on the substrate
and its evaporation-induced orthoradial shrinkage, the foot rapidly experiences 
regularly-spaced radial cracks (see Figure \ref{fracture1}).
The periodicity of the fractures in such evaporating colloidal solutions has
turned out to originate in the interplay between the stress relaxation induced
by the crack and the stress enhancement yielded by the water drainage from the
inner liquid to the crack surface \cite{Allain}.

But unlike the case of mineral colloids, the subsequent evolution of our crack
pattern is characterized by the stop of a growing number of fractures, always
preceded by a right angle turn.
In the same time, purely orthoradial cracks appear (see Figure \ref{fracture2}).
Finally all radial fractures cease to propagate simultaneously and,
depending on the protein concentration, two scenarii may occur in the central
zone: Small-scaled disordered fracturation or dendritic crystallisation
(see Figure \ref{fracture3} and \cite{AnnarelliCE}).
Respective behaviours as a function of the BSA concentration are gathered in
Table \ref{values}.

\section{Dynamics}
Before studying the fracturation, we concentrate on the evaporation dynamics.
To evaluate our dessication time $\tau_D=\frac{R_0}{v_E}$ with
$v_E=-\frac{1}{S}\frac{dV}{dt}$ water flux per unit surface ($R_0$ initial base
radius, $S$ surface and $V$ volume of the drop), we have performed measurements
of the evolution of the drying drop height $h$ with time $t$ (see Figure
\ref{height}).
Considering a constant base radius and a spherical cap shape and choosing for
$S$ the overall free surface of the drop, including the foot where evaporation
still takes place \cite{ParisseL}, the dessication time can simply be written
$\tau_D=-\frac{2R_0}{(\frac{dh}{dt})_{t=0}}=1.3\times 10^4$ s.

Furthermore, weighing measurements shown in Figure \ref{mass} have brought an
additional evaluation of the dessication time with the above-mentionned
assumptions: $\tau_D=-\frac{\rho R_0 S_0}{(\frac{dm}{dt})_{t=0}}$
with $S_0$ the initial air-drop surface, $m$ the mass of the drop and $\rho$
the density of the solution, determined by picnometry.
The obtained numerical value $\tau_D=1.5\times 10^4$ s fits with the previous
determination.
As can be stated in Figure \ref{mass}, evaporation dynamics compares tightly
in pure water and protein solution.
Indeed, the volume fraction of the protein keeping small, $\tau_D$ is quite
independent of the BSA concentration.

Then we have focussed on the properties of the solid deposit.
Such protein solutions are known to gel during dessication, helped by
the increasing ionic strength \cite{Magdassi}.
To check the physical nature of the obtained biopolymer gel, we placed the slide
in water and observed complete redilution of the deposit.
Measurements of adsorption spectra and surface tension of the sol before and
after the gelation and redilution process give identical results, proof of
the non-chemical nature of the bonds in the solid.

To evidence the sol-gel transition kinetics, we have carried out electrical
conductivity measurements of the evaporating drop by placing electrodes on the
slide, at two edge points of the drop.
As can be seen on Figure \ref{elec}, the progressive hindering of ionic
conduction through the liquid by the connection of the protein molecules gives
rise to a decrease in the current.
We have taken as gelation time $\tau_G$ the elapsed time at the achievement
of the low-conductivity plateau.
$\tau_G$ is found to decrease smoothly when concentration rises
(see Table \ref{values}).

Finally the cracking characteristic time at which the first fracture is detected
has been measured and found quite independent of concentration:
$\tau_C=8\times 10^2$ s.

\section{Initial cracking}
We now point out the fracturing behaviour of the gel.
The fact that the central part of the drop remains liquid and the three phase
line anchored while the foot starts shrinking results in orthoradial stress,
then in radial cracks.
In a quantitative theory of the 1D fracturing behaviour of a mineral colloidal
suspension confined in a Hele-Shaw cell, Limat \& Allain have found that the
crack spacing $\lambda$ should be roughly proportional to the cell thickness
\cite{Allain}.
We have implemented this model for our open geometry, considering that the same
stress field exists in our drop foot as in one longitudinal half of their slab,
leading to an equivalent cell thickness of twice the foot height.
The theoretical and experimental initial crack spacings $\lambda_0$ are
reported in Table \ref{values}.
Considering the degree of approximation of the model and the difficulty for the
physical parameters evaluation, agreement is found satisfactory.

The foot heights listed in Table \ref{values} have been measured with a
holographic interferometry device, thanks to an original geometrical analysis
\cite{AnnarelliT}.
With this technique, a reference hologram of the studied surface is first
recorded with the use of two beams (reference and object), the incident angle of
the object beam being $\alpha$.
This hologram contains information on the amplitude and phase of the
reflected object beam, enabling a 3D reconstruction of the surface.
Then a mirror is tilted, changing the incident angle of the object beam to
$\alpha+\Delta\alpha$ and a new hologram is recorded.
The interference of the two holograms results in a fringe pattern, an example of
which is shown in Figure \ref{holo}.
We have shown in \cite{AnnarelliT} that
flat surfaces lead to fringes parallel to the tilt axis, whereas any defect of
height $m$ in the surface creates a fringe shift $p$ with 
$p=m\times \tan{(\alpha+\Delta\alpha)}$.
The deposit topography has therefore been derived from the measurement of the
fringes deviation on the interference pattern caused by a tilt
$\Delta\alpha=0.5^\circ$ from the incident angle $\alpha=80^\circ$.
Validation of the results have been provided by complementary profilometer
measurements.
This method has moreover permitted to check that the flatness of the slide was
better than 2 $\mu$m over the distances investigated here.

\section{Cracking evolution}
\begin{figure}
\includegraphics[width=\linewidth]{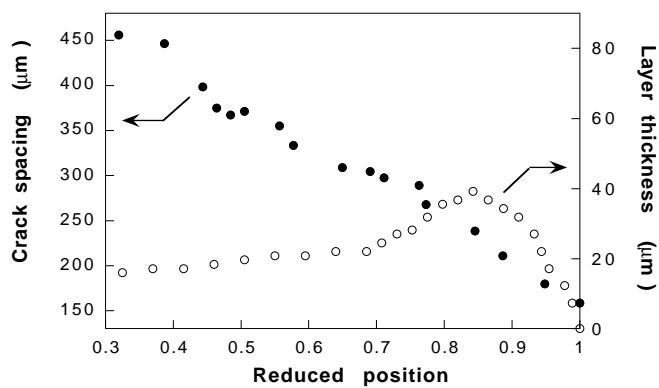}
\caption{Layer thickness $e$ and crack spacing $\lambda$ as a function of
the reduced distance $r/R_0$ from the center of a dried 40 g.l$^{-1}$ drop.}
\label{lambdae40}
\end{figure}

\begin{figure}
\includegraphics[width=\linewidth]{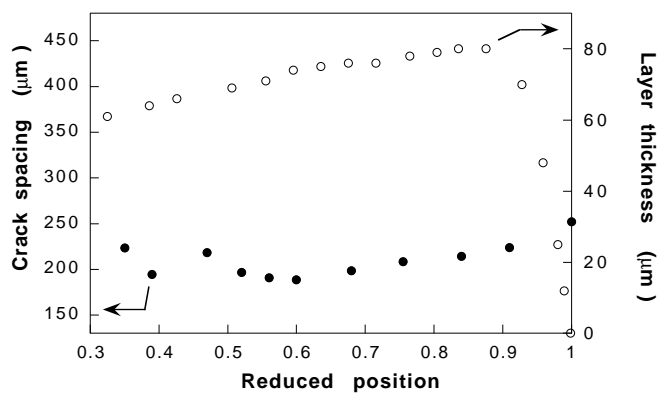}
\caption{Idem as Figure \ref{lambdae40} (same scales) for a dried
60 g.l$^{-1}$ drop.}
\label{lambdae60}
\end{figure}

After the initial breaking, number of cracks merge during propagation and the
mean crack spacing $\lambda$ may increase.
In the vicinity of a crack, the stress is purely parallel to its edges, so when
a radial fracture tends to stop, it experiences a turn to join an adjacent one
with a right angle.
The evolution of the mean crack spacing
$\lambda$ (taken as the ratio of a perimeter to the number of
fractures crossing it) when going from the center to the edge of the dried
drop is shown in Figures \ref{lambdae40} and \ref{lambdae60} with same scales
for 40 and 60 g.l$^{-1}$ BSA concentrations.

As has been qualitatively \cite{Groisman} and quantitatively (see above)
demonstrated in different configurations, the crack spacing should be
roughly proportional to the layer thickness.
We have therefore used our holointerferometry measurements
to get informations about the thickness of the deposit along the radius of the
dried drop.
Results are represented in Figures \ref{lambdae40} for the 40 g.l$^{-1}$ case
and \ref{lambdae60} for the 60 g.l$^{-1}$ case and give a clear view of the
fact that the link between crack vanishing and layer thickness is different
from what could be expected.

In fact, if we evaluate the characteristic time of diffusion of water in the gel
between two cracks during the initial stage, the order of magnitude is found 
greater than 10 minutes, i.e., comparable to the whole fracturation time
($\sim$15 min.).
So after the initial cracking, evaporation ceases to constitute a leading
feature of the phenomenon.
Therefore we must now look back at the involved kinetics to assess the governing
mechanisms.

\section{Crack pattern}
For an identical time hierarchy ($\tau_C<\tau_G<\tau_D$), Pauch\-ard
\textit{et al.} observed disordered crack patterns during the evaporation of an
aqueous silica sol deposited drop, caused by a buckling instability
\cite{Pauchard}.
This one consists of a distortion  of the drop shape caused by the gelation of
the surface when the inner sol is still liquid.
We do not face this morphological behaviour, the central zone solidifying
progressively through the propagation of the gel front.

Before ascribing this discrepancy to the respective natures
of the mineral and biopolymer gels, the influence of the geometrical parameters
has to be investigated.
Showing an axial symmetry, only two dimensions have to be probed.
First the drop volume has been varied from 2 to 25 mm$^3$ with solutions of
concentrations 20 to 60 g.l$^{-1}$.
Then, for identical concentrations,
the contact angle has been changed by testing several substrates of
biomedical interest, the wettability characteristics of which had already been
studied \cite{AnnarelliJCIS}: 
nitrogen implanted (59$^\circ$) and virgin (62$^\circ$) titanium based alloy
TA6V,
GUR 415 (65$^\circ$) and Chirulen (67$^\circ$) ultra high molecular weight
polyethylenes,
sintered alumina (66$^\circ$) and zircona (67$^\circ$),
stainless steel 316L (68$^\circ$), and
a chromium-cobalt alloy (71$^\circ$).
Additionally, silane-coated glass ($>$90$^\circ$) has been tested.
Whereas the fractures penetration depth variation with concentration and
contact angle show interesting features ---which will be studied elsewhere---,
in all cases the drop morphology and cracking evolution are
similar to the case of the 15 mm$^3$ drop deposited on glass.
One only difference can be noticed for volumes smaller or equal to 5 mm$^3$
and silane-coated glass
where all initial radial fractures stop propagating at the same time
instead of hierarchically due to the small base radius of the drop.

So the universal radial nature of the crack pattern (instead of a disordered
nature for a mineral gel) seems to be independent of geometrical considerations
and can reasonably be attributed
to the physical nature of the biopolymer gel, the latter exhibiting smoother
mechanical properties than chemical gels.
Its low tensile strength in the gelling central zone should thereby result
in an overall stress in the gel front keeping an orthoradial direction.

\section{Mean crack spacing evolution}
The dessication kinetics being non-critical after the initial cracking,
information is now transmitted elastically between cracks, through the flux
of displacement field, as pointed out in impact fracture experiments
\cite{Inaoka}, which is a much faster mode.
Thus we face a pattern selection system comparable to the one
encountered in thermal crack growth experiments \cite{Ronsin}: The release
of elastic energy during the propagation controls the crack spacing.

Indeed for the propagation of a crack to be possible, the elastic energy $E_e$
liberated by the breaking must at least equal the surface energy $E_s$ required
by the two formed lips (Griffith's criterion).
We are considering here as a first approximation purely brittle breaking,
dissipative effects (plasticity, friction) being neglected.
During the progression $\delta l$ of a fracture, $E_e$ is
proportional to the relaxed volume $e\times\lambda\times\delta l$ and $E_s$ to
the created surface area $e\times\delta l$.
So the fulfilment of the $E_e=E_s$ condition does not imply, in the case of a
change of $e$, a modification of the crack spacing.
Therefore the evolution of $\lambda$ when approaching the center of the drop,
if any,
should be related to a change in the only other varying factor along the
fracture path: the stress field.

For example, in the 40 g.l$^{-1}$ case, $\lambda$ evolves roughly as
$r^{-\frac{1}{2}}$ ($r$ distance from the center
of the drop), so linear elasticity yields a change of the stress $\sigma$
proportional to $\lambda^{-\frac{1}{2}}$ so to $r^{\frac{1}{4}}$.
This slow decrease of $\sigma$ when going from the edge to the center of the
drop has to be related to the decrease of the average stress with time: Cracks
meet during their propagation a gel more and more relaxed by the continuous
fracturation \cite{Skjeltrop}.
In the 60 g.l$^{-1}$ case, $\lambda$ exhibits no actual evolution, which should
stem from a constant stress during the cracking process.
This constancy of the stress is likely to be a consequence of the large
and quite uniform layer thickness, which hinders relief mechanisms.

So the dynamics of the mean crack spacing has revealed itself to be
actually related to the
deposit thickness, but not in the expected way: It is rather inversely
proportional instead of proportional.
Indeed, dessication playing no more role, only cracking influences the cracking
geometry through stress relief, which is made easier by decreasing layer
thicknesses.

\section{Conclusion}
We have investigated the evaporating, gelling and cracking behaviour of a
deposited drop of protein solution.
The initial stage is characterized by the appearance of regularly-spaced cracks
at the edge of the gelling drop.
This ordered patterning was known to originate in the interplay between the
water diffusion towards the crack surface and the elastic relaxation due to
the fracture.
The general radial patterning was found to be a corollary of the soft nature of
the BSA gel.
Then a transition from this evaporation-controlled behaviour to a purely
elastic one occurs, where the change of the crack spacing has turned out
to be a consequence of the shrinkage stress evolution with time.

This first study of the fracturing behaviour of a drying biological liquid
has shown similarities with other complex fluids.
But discrepancies have also been emphasized and an investigation of the specific
mechanical properties of the biopolymer gel is still needed to confirm the
assumptions made on the dynamics of the fracturation.
The influence of the substrate on the cracking, especially on fracture
penetration depths, constitutes also a promising way of investigation.

\begin{acknowledgement}
We acknowledge fruitful discussions with Laurence Reyes and Christian Olagnon
and experimental help from Richard Cohen, Jean-Alain Roger and
Stella Ramos-Canut.
\end{acknowledgement}

\begin{center}
\begin{figure}
\includegraphics[angle=270,width=\linewidth]{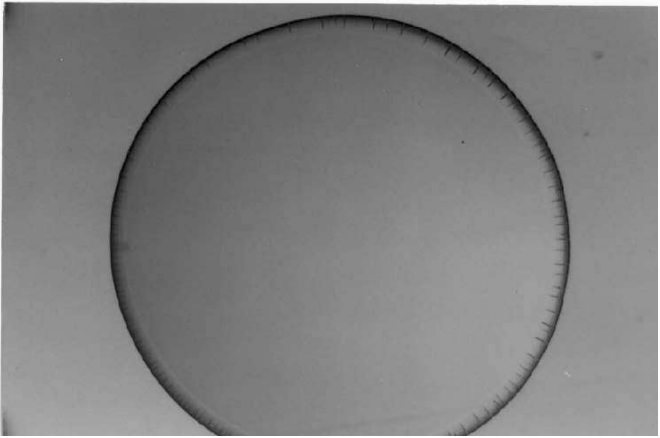}
\caption{Beginning of the foot fracturation of a drying 40 g.l$^{-1}$ solution
drop 840 s after deposit.}
\label{fracture1}
\end{figure}

\begin{figure}
\includegraphics[angle=90,width=\linewidth]{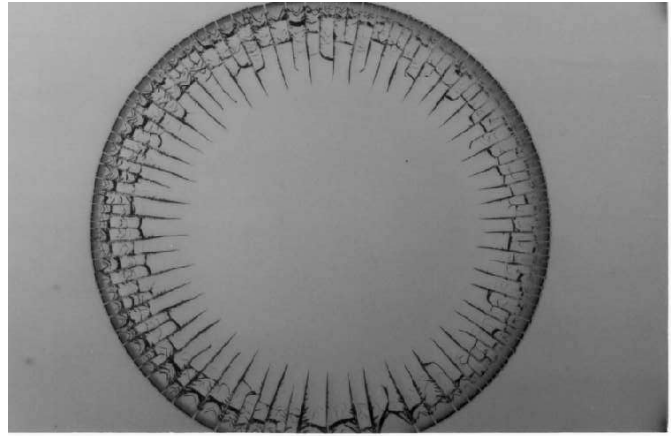}
\caption{Following of the fracturation of the drying drop of Figure 
\ref{fracture1}, 500 s later.}
\label{fracture2}
\end{figure}

\begin{figure}
\includegraphics[angle=90,width=\linewidth]{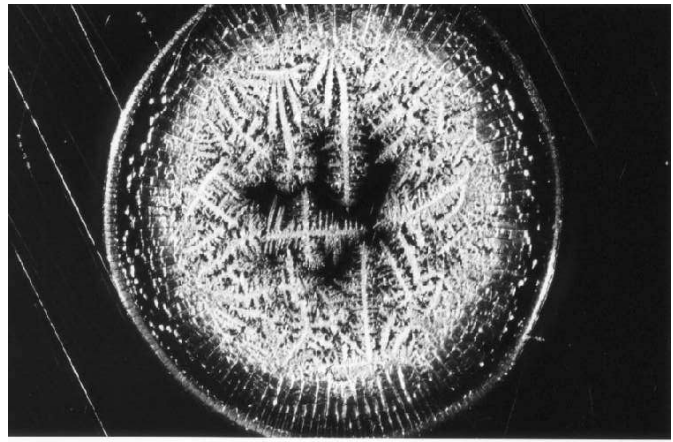}
\caption{Dendritic crystallisation of the central zone of a dried 10 g.l$^{-1}$
solution drop.}
\label{fracture3}
\end{figure}

\begin{figure}
\includegraphics[width=\linewidth]{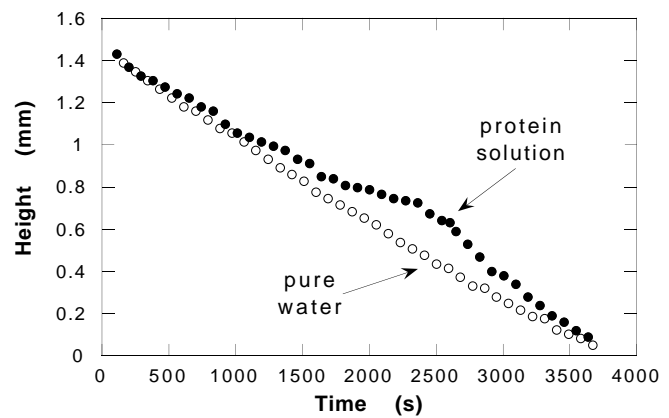}
\caption{Time evolution of a drying 40 g.l$^{-1}$ solution drop height.
As a matter of comparison the values for a pure water drop are also displayed.}
\label{height}
\end{figure}

\begin{figure}
\includegraphics[width=\linewidth]{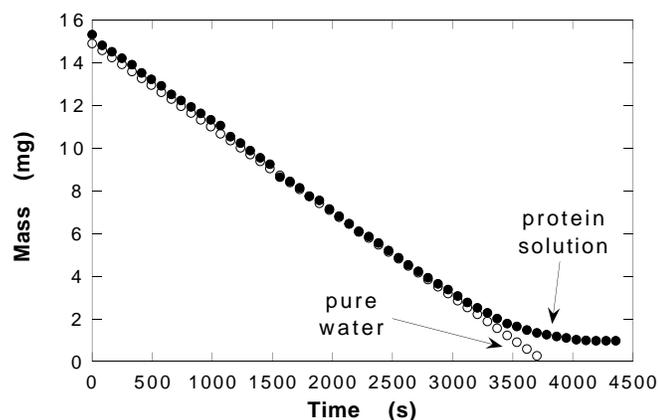}
\caption{Time evolution of a drying 40 g.l$^{-1}$ solution drop mass.
Values for pure water are also displayed.}
\label{mass}
\end{figure}

\begin{figure}
\includegraphics[width=\linewidth]{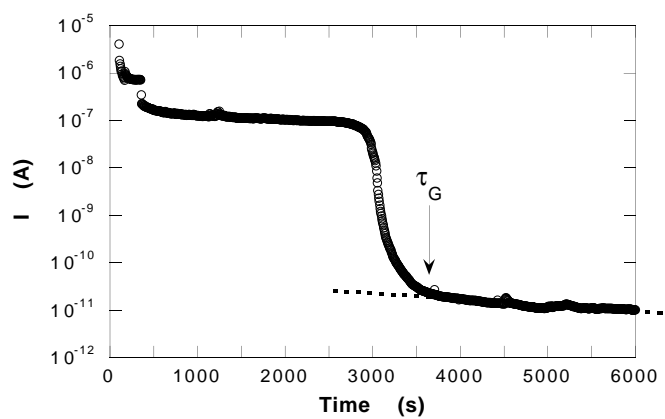}
\caption{Time evolution of the electric current between two edge points of the
drying 40 g.l$^{-1}$ solution drop.}
\label{elec}
\end{figure}

\begin{figure}
\includegraphics[width=0.7\linewidth]{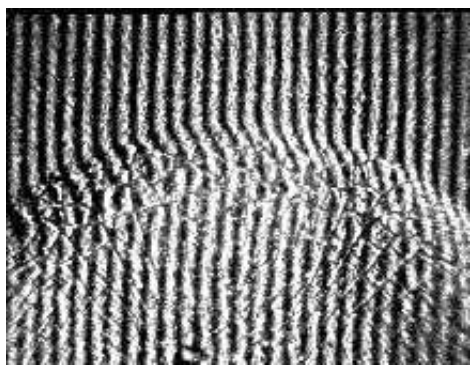}
\caption{Holointerferogram of a dried 40 g.l$^{-1}$ drop deposit.}
\label{holo}
\end{figure}
\end{center}

\begin{table}
\begin{tabular}{c || c | c | c | c | l}
$c$ & foot height & $\tau_G$ & $\lambda_0^{exp}$ & $\lambda_0^{theo}$ & final behaviour\\
(g.l$^{-1}$) & ($\mu$m) & (10$^3$ s) & ($\mu$m) & ($\mu$m) & \\
\hline\hline
20 & 28 & 3.8 & 100 & 124 & crystallisation\\
30 & 31 & 3.7 & 120 & 135 & crystallisation\\
40 & 36 & 3.6 & 160 & 154 & fracturation\\
60 & 81 & 3.4 & 250 & 309 & fracturation\\
\end{tabular}
\caption{Foot height, gelation characteristic time, experimental value of the
initial crack spacing, theoretical value of the initial crack spacing, final
behaviour of the deposit center as a function of BSA concentration.}
\label{values}
\end{table}

\end{document}